\documentclass[11pt]{article}
\usepackage{amsmath,amssymb,amsfonts,graphicx,listings}
\usepackage[left=1in,right=1in,bottom=1in]{geometry}
\title{A Validation \\of\\Causal Dynamical Triangulations}
\author{Rajesh Kommu\footnote{\textit{kommu@physics.ucdavis.edu}}\\
	{\small\it Department of Physics}\\
       {\small\it University of California}\\
       {\small\it Davis, CA 95616}\\{\small\it USA}}

\newcommand{\mym}[1]{$\displaystyle #1$}

\newcommand{\code}[1]{\texttt{#1}}

\begin{document}
\maketitle
\begin{abstract}
The Causal Dynamical Triangulation (CDT) approach to quantum gravity is a 
lattice approximation to the gravitational path integral. Developed by 
Ambj\o{}rn, Jurkiewicz and Loll, it has yielded some important results, notably
the emergence of classical spacetime and short scale dimensional reduction. 
However, virtually all the results reported so far have been based on a single computer 
code. In this paper we present the first completely independent verification of the 
CDT algorithm, and report the successful reproduction of the emergence of
classical spacetime and smooth reduction in the spectral dimension of the 
2+1 and 3+1 dimensional spacetimes.
\end{abstract}


\section{Introduction}\label{sec:intro}
The Causal Dynamical Triangulations (CDT) approach to quantum gravity has
emerged as a breath of fresh air in the effort to quantize gravity. In CDT,
developed by Ambj\o{}rn, Jurkiewicz and Loll 
\cite{C0,C4,C1,C2}, we have an attempt at
quantizing gravity that relies solely on ideas and techniques that have been 
known to physicists for quite some time --- path integrals, causality, 
simplicial manifolds, finite-size scaling effects, etc. 
The formulation of CDT, while being extremely straightforward, has yielded 
some important results, such as the \textit{emergence} of classical 
spacetime and dimensional reduction of spacetimes at short distances.

So far, however, all the results that have been reported for CDT have either 
come from the original group headed by Ambj\o{}rn, Jurkiewicz and Loll or from
groups that have used the code or data sets from the original group. In 
this paper, we present the first independent verification of the CDT algorithm,
using code developed from scratch in a completely independent fashion. We have 
been able to reproduce the important results reported for CDT and
set the stage for further results, which will be reported elsewhere
\cite{Cooperman,Sachs,Zulkowski}. 

The outline of this paper is as follows. In section \ref{sec:cdt} we present
a brief overview of the Causal Dynamical Triangulation approach. Readers 
interested in greater detail should refer to \cite{C4} and \cite{C1}. In 
section \ref{sec:num-impl} we describe our CDT implementation, at a 
fairly high level. Finally, in section \ref{sec:results}, we present the 
results from our simulation.
\section{Causal Dynamical Triangulations}\label{sec:cdt}
CDT is based on the gravitational path integral
\begin{equation}\label{eqn:grav-path-integral}
G(\mathbf{g_i},\mathbf{g_f};t_i,t_f)=\int\mathcal{D}[g]e^{iS_{EH}[g]}
\end{equation}
where
\begin{equation}\label{eqn:Einstein-Hilbert-action}
S_{EH}[g]=\frac{1}{16\pi G}\int d^nx\sqrt{-g}(R-2\Lambda)
\end{equation}
is the Einstein-Hilbert action. Here \mym{\mathbf{g_i}} and \mym{\mathbf{g_f}}
are the spatial metrics on the initial and final slice and
\mym{g} represents all possible (spacetime) metrics that satisfy the initial 
and final conditions. By all possible metrics, we mean metrics that are not 
related to each other by diffeomorphisms. 
To define the action in the path integral, CDT uses Regge calculus \cite{00}
as its starting point. The basic idea of Regge calculus is to approximate a
smooth manifold (for example spacetime) with a piecewise linear manifold, 
with curvature restricted to subspaces of co-dimension two.

The geometrical objects used to approximate $d$-dimensional manifolds are
$d$-simplices, which consist of $(d-1)$-dimensional faces, 
$(d-2)$-dimensional hinges or ``bones'' (on which the curvature
of the manifold resides), and so on down to 0-dimensional simplices
(points). A $d$-simplex has $d+1$ points.

For manifolds without a boundary, a discrete version of the Einstein-Hilbert 
action can be formulated using \cite{00}
\begin{eqnarray}\label{eqn:regge-results}
\int d^nx\sqrt{-g}R&\rightarrow& 2\sum_{hinges~h}V_h\delta_h\\
\int d^nx\sqrt{-g}&\rightarrow& \sum_{d}V_d
\end{eqnarray}
to get
\begin{equation}\label{eqn:regge-action}
S_R=\frac{1}{8\pi G}\sum_{h}V_h\delta_h-\frac{\Lambda}{8\pi G}\sum_{d}V_d
\end{equation}
where $V_h$ is the volume of a hinge, $V_d$ is the volume of a top level
d-simplex, and $\delta_h$ is the deficit angle at $h$.

CDT assumes globally hyperbolic spacetimes. The existence of Cauchy hypersurfaces
enables us to use $d$-simplices that span adjacent spatial slices as the building
blocks for our $d$-dimensional spacetimes. A time-like $(p,q)$ $d$-simplex, where 
$p+q=d+1$, has $p$ points on the lower time slice and $q$ points the adjacent higher 
time slice. For a space-like simplex, all points are on the same spatial slice.
No space-like $d$-simplices are present in $d$-dimensional spacetimes. But both
space-like and time-like types of lower dimensional simplices are present in
the spacetime. The simplicial building blocks in CDT have an inherent causal structure
built into them. To preserve this causal structure, CDT does not allow topology
changes of the spatial slices. 

The lengths of the space-like and time-like links (1-simplices) are defined as
\begin{equation}\label{eqn:sl-and-tl}
l^2_{space}=a^2, l^2_{time}=-\alpha a^2, \alpha > 0
\end{equation}
where $a$ is the lattice spacing and $\alpha$ is the asymmetry parameter.
In 2+1 dimensions, we have three types of 3-simplices, denoted by $(1,3)$,
$(2,2)$ and $(3,1)$. In 3+1 dimensions, we have four types of 4-simplices, 
denoted by $(1,4), (2,3), (3,2)~\text{and}~(4,1)$.
The volume of each type of simplex and the dihedral angles subtended between
the faces of these simplices are given in \cite{C4}. Using these, we can 
formulate the Lorentzian version of the Regge action in $d=3,4$ as
\begin{equation}\label{eqn:cdt-analogue-regge-action}
\begin{split}
S=\frac{1}{8\pi G}\sum_{\substack{space-like\\hinges,h}}Vol(h)\frac{1}{i}\biggl(2\pi -
\sum_{\substack{d-simplices\\at~h}}\Theta\biggr) \\+
\frac{1}{8\pi G}\sum_{\substack{time-like\\hinges,h}}Vol(h)\biggl(2\pi -
\sum_{\substack{d-simplices\\at~h}}\Theta\biggr) \\-
\frac{\Lambda}{8\pi G}\sum_{\substack{time-like\\d-simplices,d}}Vol(d)
\end{split}
\end{equation}
where $\Theta$ is the dihedral angle subtended at the hinge $h$.  
Equation (\ref{eqn:cdt-analogue-regge-action}) is a reformulation of 
(\ref{eqn:regge-action}) with space-like and time-like simplices 
distinguished.

In CDT, Wick rotation is defined as \cite{C4}
\begin{equation}\label{eqn:wick-rot}
\alpha \mapsto -\alpha, \alpha > \frac{1}{2} (d=3), \alpha > \frac{7}{12} (d=4)
\end{equation} 
This leads to a pure imaginary value for the action. The gravitational path 
integral (\ref{eqn:grav-path-integral}) in CDT is replaced by a sum over 
distinct triangulations 
\begin{equation}\label{eqn:sum-over-triangulations}
\int \mathcal{D}[g]e^{iS} \rightarrow \sum_T\frac{1}{C(T)}e^{iS(T)}
\end{equation}
where $C(T)$ is the order of the automorphism group of the triangulation.

The Wick rotated action\footnote{Equations (\ref{eqn:2p1-wick-rotated-action})
and (\ref{eqn:2p1-kappa0-kappa3}) are derived for $\alpha=-1$. For 
$\alpha\neq -1$ the action will have terms involving $N_0,N_3^{(3,1)}$ and 
$N_3^{(2,1)}$, where $N_3^{(3,1)}$ and $N_3^{(2,2)}$ are the number of (3,1) and 
(2,2) 3-simplices. However, in 2+1 dimensions, there are only two independent 
bulk degrees of freedom, so the action still takes the form 
(\ref{eqn:2p1-wick-rotated-action}), although with 
(\ref{eqn:2p1-kappa0-kappa3}) altered. See \cite{C4} for details.} for a 
triangulation in 2+1 dimensions is given by
\begin{equation}\label{eqn:2p1-wick-rotated-action}
S^{(3)}=i(-\kappa_0N_0+\kappa_3N_3)
\end{equation}
where
\begin{equation}\label{eqn:2p1-kappa0-kappa3}
\kappa_0=\frac{1}{4G},\kappa_3=\frac{\Lambda}{48\sqrt{2}\pi G}+
\frac{1}{4G}\biggl(\frac{3}{\pi}\cos^{-1}(1/3)-1\biggr)
\end{equation}
and $N_0,N_3$ are the number of 0-simplices and 3-simplices, respectively. 
Equation (\ref{eqn:2p1-wick-rotated-action}) can be derived from 
(\ref{eqn:cdt-analogue-regge-action}) by substituting the 
appropriate expressions for the simplex volumes and dihedral angles, using
topological identities for (2+1)-dimensional Lorentzian triangulations \cite{C4}
and then performing a Wick rotation (\ref{eqn:wick-rot}). The topological
identities eliminate all but two of the bulk variables $N_0\cdots N_d$.

Substituting the Wick rotated action in (\ref{eqn:sum-over-triangulations})
gives us the partition function:
\begin{equation}\label{eqn:2p1-partition-function}
Z=\sum_T \exp(\kappa_0N_0-\kappa_3N_3)
\end{equation}

A similar procedure in 3+1 dimensions leads to the Wick rotated action
\begin{equation}\label{eqn:3p1-wick-rotated-action}
S^{(4)}=i\biggl(-(\kappa_0+6\Delta)N_0+\kappa_4N_4+\Delta(2N_4^{(4,1)}+N_4^{(3,2)})\biggr)
\end{equation}
and the corresponding partition function
\begin{equation}\label{eqn:3p1-partition-function}
Z=\sum_T \exp\biggl((\kappa_0+6\Delta)N_0-
\kappa_4N_4-\Delta(2N_4^{(4,1)}+N_4^{(3,2)})\biggr)
\end{equation}
where
\begin{equation}\label{eqn:3p1-kappa0-kappa4}
\kappa_0=\frac{\sqrt{3}}{8G},\kappa_4=\frac{\Lambda\sqrt{5}}{768\pi G}+
\frac{\sqrt{3}}{8G}\biggl(\frac{5}{2\pi}\cos^{-1}(1/4)- 1\biggr)
\end{equation}
and $\Delta$ is a complicated function that depends on $\alpha$, with 
$\alpha=1$ corresponding to $\Delta=0$. $N_4^{(4,1)}$ is the number of $(4,1)$ 
plus $(1,4)$ simplices, and $N_4^{(3,2)}$ is the number of $(3,2)$ plus $(2,3)$ 
simplices. 

To generate the distinct triangulations required for evaluating the partition
function and the expectation value of any observable, the CDT approach uses a 
set of ergodic moves. An ergodic move applied to a triangulation yields 
another distinct triangulation, one that is not related to the first one by a 
diffeomorphism. Ergodicity ensures that we can go from one triangulation to 
any other triangulation by repeated application of moves. Ergodic moves also 
preserve the topology of the manifold, as well as 
the simplicial manifold property. The moves are described in detail in 
\cite{C4}.

\section{Numerical Implementation}\label{sec:num-impl}
In our numerical implementation, each $d$-simplex is identified by a unique 
ID and is stored as a list (containing sublists) of points
\begin{lstlisting}
(ty lo hi (p0 p1 . . . pd) (n0 n1 . . . nd))
\end{lstlisting}
Here \code{ty} identifies the type of the simplex \code{ty = 1,2,...,d}.  
\code{lo} and \code{hi} are the lower and higher time slices spanned by the 
simplex. The \code{pis} are the points that make up the simplex, and the
\code{nis} are the IDs of the neighboring simplices. The simplex \code{nk} is
connected on the \code{(p0 ... pd) - (pk)} face. The points are used only as unique 
labels to identify the vertices of a simplex. While we distinguish the points on the 
lower time slice from the points on the higher time slice, the ordering of the
points in any given slice is immaterial. 

With this choice for the simplex data structure, it is fairly straightforward
to implement the ergodic moves described in detail in \cite{C4}. It is essential
that the moves maintain the simplical property of the manifold. This requires
that two simplices should be connected across a single face and that a 
$(d-1)$-simplex should be shared between exactly two $d$-simplices.

To generate the distinct triangulations using the ergodic moves, we apply the
Metropolis algorithm \cite{8I}. The Metropolis algorithm uses a Markov chain 
that, upon reaching equilibrium, satisfies the following detailed balance 
condition  
\begin{equation}\label{eqn:detailed-balance-1}
P(T_1)W(T_1,T_2)=P(T_2)W(T_2,T_1)
\end{equation}
where $P(T_i)$ is the probability of being in a particular triangulated geometry
$T_i$ and $W(T_i,T_j)$ is the transition probability of going from triangulation 
$T_i$ to triangulation $T_j$. The probability of being in a particular 
triangulation is proportional to the Boltzmann weight $e^{-S_E}$. The detailed
balance condition in this case can be written as:
\begin{equation}\label{eqn:detailed-balance-2}
\frac{W(T_1,T_2)}{W(T_2,T_1)}=\frac{e^{-S_E(T_2)}}{e^{-S_E(T_1)}}=e^{-\Delta S_E}
\end{equation}
The simplest choice for the transition probability $W$ that satisfies this
condition is
\begin{equation}\label{eqn:w-choice}
W(T_1,T_2)=
\begin{cases}
e^{-\Delta S_E}, & \text{if }\Delta S_E>0 \\
1, & \text{if }\Delta S_E\leq0
\end{cases}
\end{equation}
Thus a move that increases the action may be accepted or rejected, while moves
that decrease the action or leave the action unchanged are always accepted.

We initialize the simulation by specifying the number of time slices $T$ and the 
total volume of the spacetime. The initialization process starts by triangulating
each spatial slice as a $(d-1)$-sphere. The $t=0$ and the $t=T$ slices are identified 
to enforce the periodic boundary conditions in time. This is done for technical 
convenience, rather than due to any physical argument. Next, the triangulations 
on adjacent slices are connected. We then apply the volume increasing subset of
the ergodic moves until the total volume reaches the desired target volume. 

Once the simulation has been initialized, we have to run simulation for about
100,000 sweeps to thermalize the spacetimes. A sweep is defined as $N_d$ attempted moves, 
where $N_d$ is number of $d$-simplices in the spacetime.
Thermalization ensures that the system is sufficiently randomized, and has moved 
away from the highly structured initial conditions. Once thermalization is completed, 
we start saving spacetimes every 100 sweeps, to build our ensembles.

\section{Results}\label{sec:results}

\subsection{Phase Structure}
\begin{figure}[htp]
\centering
\includegraphics[scale=0.5]{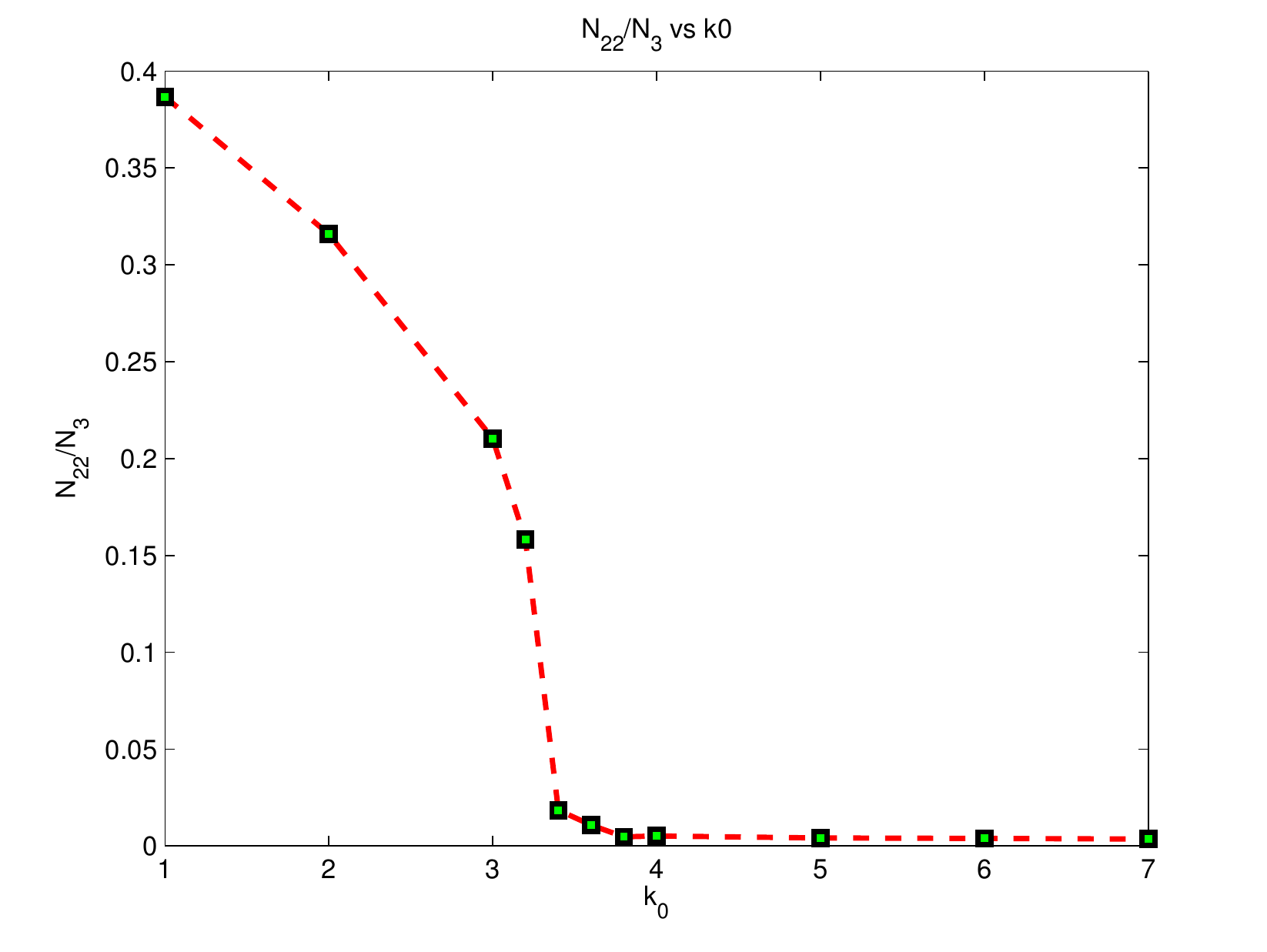}
\caption{$\frac{N_{22}}{N_3}$ versus $\kappa_0$}\label{fig:op_vs_kzero}
\end{figure}

In 2+1 dimensions, the phase structure of the CDT model can be explored by 
tracking the dependence of an order parameter on coupling constants. The 
order parameter selected is the ratio of the number of \mym{(2,2)} 3-simplices
to the total number of 3-simplices in the spacetime:
\begin{equation}
\tau=\frac{N_{22}}{N_3}=\frac{N_{22}}{N_{13}+N_{22}+N_{31}}
\end{equation}
The dependence of \mym{\tau} on the coupling constant is shown in figure 
\ref{fig:op_vs_kzero}. 

\begin{figure}[htp]
\centering
\includegraphics[scale=0.5]{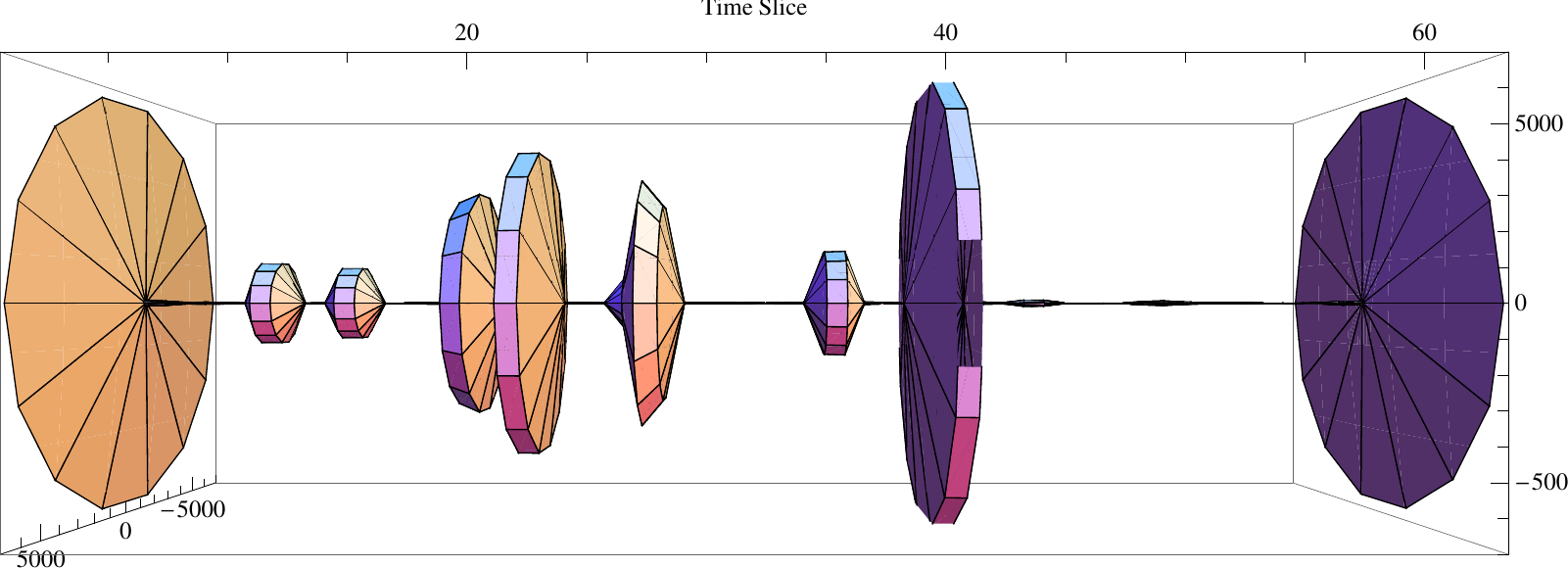}
\caption{A snapshot of a (2+1)-dimensional spacetime in the decoupled phase.}
\label{fig:decoupled_phase}
\end{figure}

Figure \ref{fig:op_vs_kzero} shows a phase transition, at 
\mym{\kappa_0^c\approx3.3}. In \cite{C1}, this phase transition was reported at a value of
\mym{\kappa_0^c\approx6.6}. The two results differ by a factor of two, for reasons we
do not understand, but the qualitative results are identical.  For \mym{\kappa_0>\kappa_0^c}, 
we observe that the spatial volumes of successive slices is not strongly correlated --- spacetime 
has effectively decoupled into disconnected spatial slices. This is shown in 
figure \ref{fig:decoupled_phase}, 
where we have a taken a snapshot of a simulation in this decoupled phase.

\begin{figure}[htp]
\centering
\includegraphics[scale=0.5]{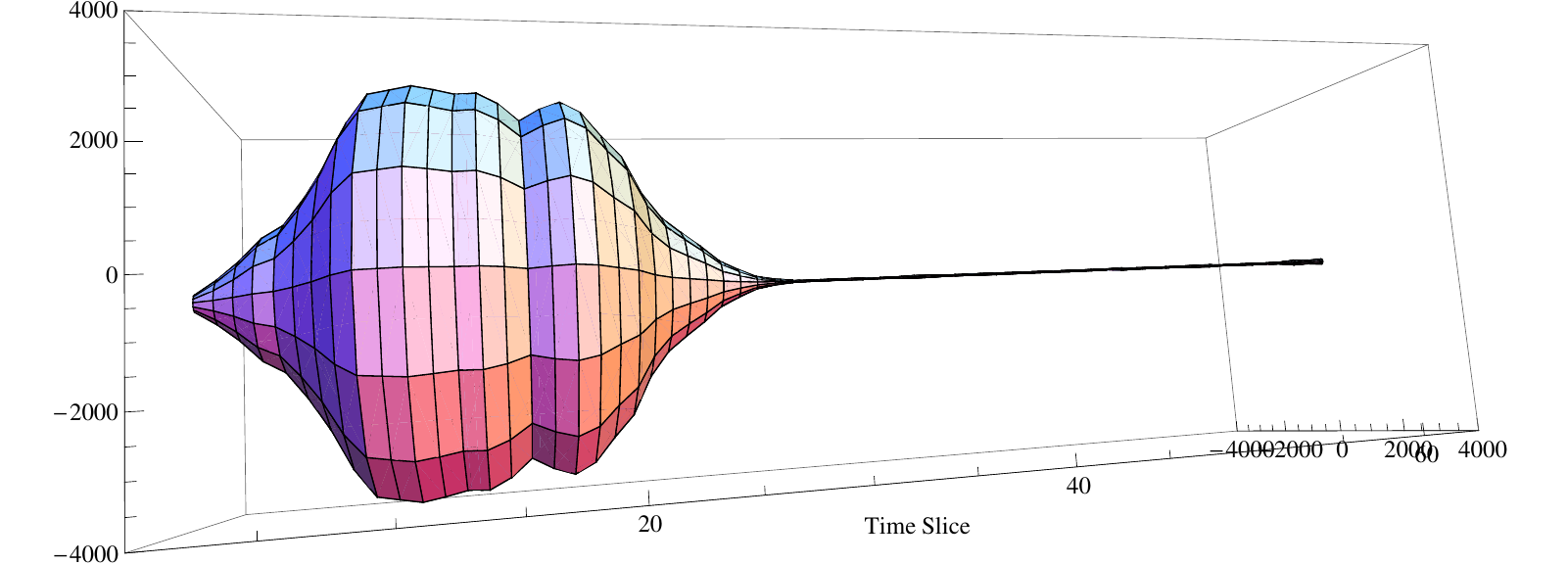}
\caption{A snapshot of a (2+1)-dimensional spacetime in the extended phase.}
\label{fig:extended-phase}
\end{figure}

A snapshot of the simulation in the other phase, characterized by 
\mym{\kappa_0<\kappa_0^c}, clearly shows the emergence of a well defined, 
extended geometry phase. This is shown in figure \ref{fig:extended-phase}.
The simulation fluctuates around this extended geometry, a geometry that 
resembles a classical universe. This classical spacetime \textit{emerges} from 
the simulation, and was not explicitly specified by us. In this phase, the 
volume of successive spatial slices is strongly correlated, and the path 
integral is dominated by metrics that are approximate solutions of the 
Einstein equations.

\begin{figure}[htp]
\centering
\includegraphics[scale=0.5]{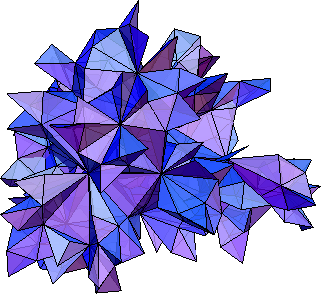}
\caption{A spatial slice in 2+1 dimensions.}
\label{fig:2p1-spatial-slice}
\end{figure}

A thick slice is defined as the portion of spacetime that is bounded by adjacent
spatial slices. Thick slices are labelled by half-integer values where, for
example, a thick slice labelled by $t=1/2$, is bounded by the spatial slices 
$t=0$ and $t=1$. The volume of a thick slice is defined as the number of 
$d$-simplices in that thick slice. 
In figures \ref{fig:decoupled_phase} and \ref{fig:extended-phase} time is 
along the horizontal axis. These figures were generated
by considering the volumes of the thick slices for half-integer values ranging
from $t=1/2$ to $t=T-1/2$ where $T$ is total number of time slices. 
We draw a circle with radius equal to the volume of the slice, and
join circles corresponding to adjacent thick slices by linear 
interpolation. The axial symmetry in these figures is therefore a visualization
artefact. Figure \ref{fig:2p1-spatial-slice} shows a typical spatial slice in
2+1 dimensions, embedded in a three dimensional space \cite{Sachs}.

\begin{figure}[htp]
\centering
\includegraphics[scale=0.5]{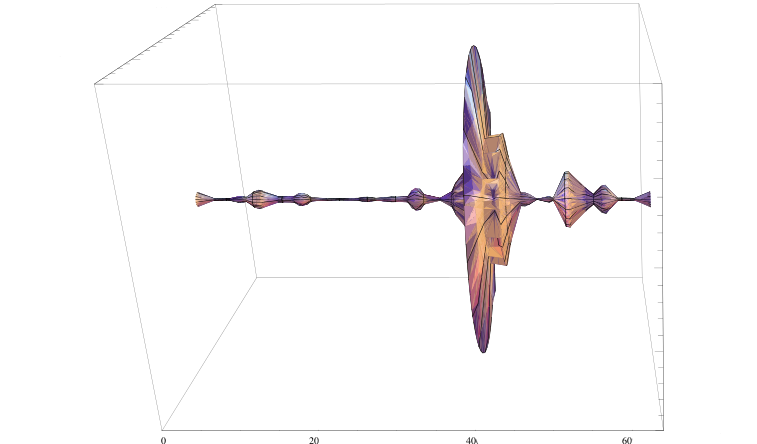}
\caption{A (3+1)-dimensional spacetime in phase A, $\kappa_0=5.0,\Delta=0.0,\kappa_4=1.01$}                                                                                                                 
\label{fig:3p1-phase-A}
\end{figure}

\begin{figure}[htp]
\centering
\includegraphics[scale=0.5]{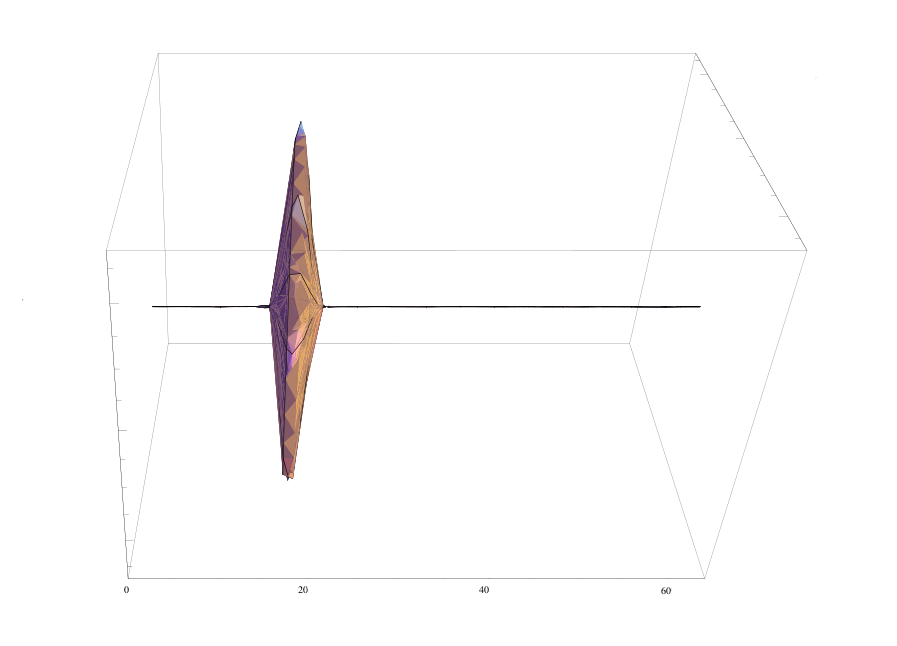}
\caption{A (3+1)-dimensional spacetime in phase B, $\kappa_0=1.6,\Delta=0.0,\kappa_4=0.65$}
\label{fig:3p1-phase-B}
\end{figure}

\begin{figure}[htp]
\centering
\includegraphics[scale=0.5]{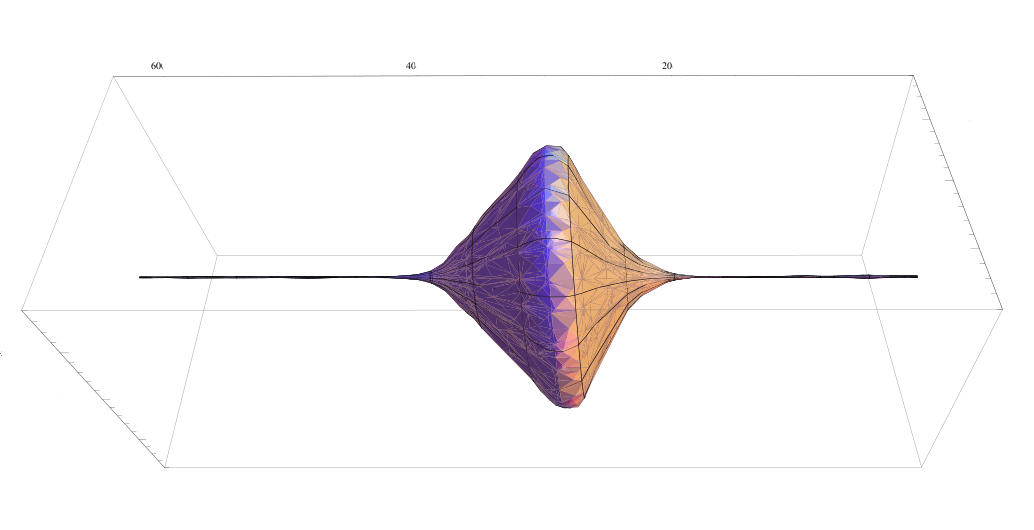}
\caption{A (3+1)-dimensional spacetime in phase C,$\kappa_0=2.4,\Delta=0.6,
\kappa_4=0.079$}
\label{fig:3p1-phase-C}
\end{figure}

In 3+1 dimensions, while we are not aware of an order parameter analogous to
the $(2+1)$-dimensional case, we do observe three
distinct phases, illustrated in figures \ref{fig:3p1-phase-A}, 
\ref{fig:3p1-phase-B} and \ref{fig:3p1-phase-C}. All three figures are for
spacetimes with $T=64$ and $N_4=81920$

Phase A is characterized by a splitting up of the spacetime in small pieces
of irregular volume, with each piece not extending beyond a few time slices. In 
Phase B, all of the spacetime volume collapses into a single piece of minimal 
extension. Both of these non-physical phases are characterized by a small value
for the ratio \mym{\frac{N_{(3,2)}}{N_4}}. 
Phase C is the extended geometric phase, where there is a strong correlation
between the volumes of successive spatial slices.

\subsection{Spectral Dimension}
One of the major contributions of the CDT approach has been the introduction
of the spectral dimension as probe of the properties of spacetime, both on 
short and long scales.

The diffusion equation for a $d$-dimensional manifold with a smooth metric
\mym{g_{\mu\nu}} is given by \cite{CD,C6}
\begin{equation}\label{eqn:diffusion-equation}
\frac{\partial}{\partial\sigma}K_g(\xi,\xi_0;\sigma)=
\Delta_gK_g(\xi,\xi_0;\sigma)
\end{equation}
where \mym{\sigma} is the diffusion time,
\mym{\Delta_g=-g^{\mu\nu}\nabla_\mu\nabla_\nu} is the Laplace operator 
corresponding to \mym{g_{\mu\nu}(\xi)} and \mym{K_g(\xi,\xi_0;\sigma)} is the 
probability density of diffusion from \mym{\xi_0} to \mym{\xi} in a time
\mym{\sigma}. The diffusion processes we care about are ones that are 
initially peaked at a point \mym{\xi_0}
\begin{equation}
K_g(\xi,\xi_0;\sigma=0)=\frac{\delta^d(\xi-\xi_0)}{\sqrt{|g|}}
\end{equation}
Instead of probability density, we measure the return probability, which is 
defined as
\begin{equation}\label{eqn:return-probability}
P_g(\sigma)=\frac{1}{V}\int_Md^d\xi\sqrt{|g|}K_g(\xi,\xi;\sigma)
\end{equation}
For a flat space with infinite volume, the solution of equation 
(\ref{eqn:diffusion-equation}) is \cite{CD,C6}
\begin{equation}\label{eqn:flat-space-heat-soln}
K_g(\xi,\xi_0;\sigma)=\frac{e^{-d_g^2(\xi,\xi_0)/4\sigma}}{(4\pi\sigma)^{d/2}}
\end{equation}
where \mym{d_g(\xi,\xi_0)} is the geodesic distance between \mym{\xi} and
\mym{\xi_0}. The return probability for the flat space case is
\begin{equation}\label{eqn:flat-space-ret-prob}
P_g(\sigma)=\frac{1}{\sigma^{d/2}}
\end{equation}
Taking the logarithmic derivative of the equation (
\ref{eqn:flat-space-ret-prob}), we get
\begin{equation}\label{eqn:ret-prob-log-der}
-2\frac{d\ln P(\sigma)}{d\ln\sigma}=d
\end{equation}
This definition of the spectral dimension can 
be extended to curved and/or finite volume spacetimes, with the addition of 
finite-size corrections, and allows one to define a generalized dimension
on a wide variety of spaces that are not smooth manifolds \cite{Carlip} 

In CDT, the spectral dimension \cite{CD,C2,CA} of the spacetime is 
measured by initiating a diffusion process from a randomly selected
simplex. A single step in a random walk is generated by moving to randomly 
selected neighbor. By initiating a large number of walks with $\sigma$ steps, 
we can calculate the probability of returning to the starting simplex. The return
probability can be then used to compute the spectral dimension.

\begin{figure}[htp]
\centering
\includegraphics[scale=0.5]{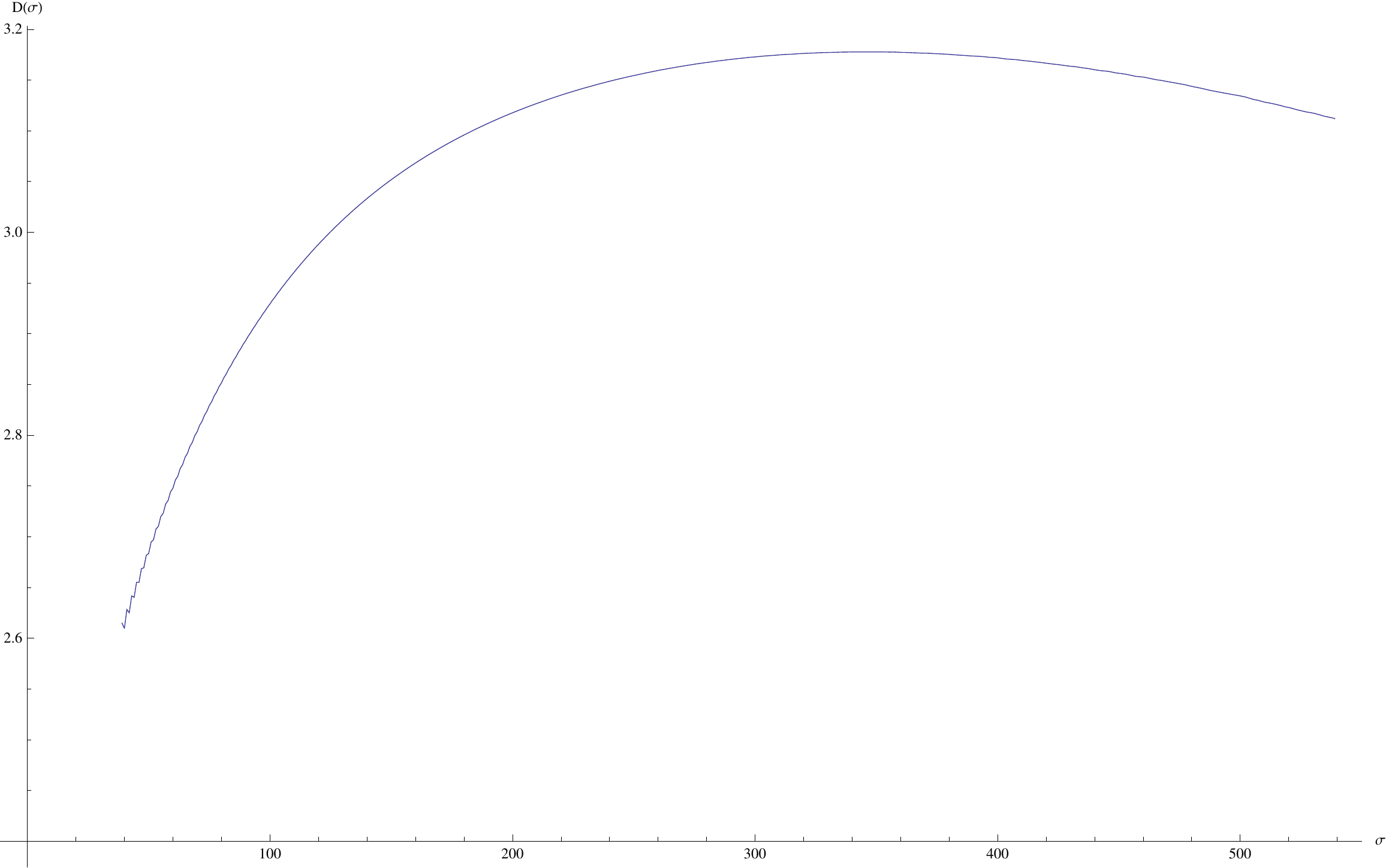}
\caption{Spectral dimension for a (2+1) spacetime vs $\sigma$, $N_3=80000,\kappa_0=1.0$}
\label{fig:2p1-spectral-dimension}
\end{figure}

The spectral dimension vs step size plot for a (2+1)-dimensional spacetime
is shown in figure \ref{fig:2p1-spectral-dimension}.
We observe that the spectral dimension demonstrates a smooth 
transition from a value of about 2.4 at short scales up to a value of about 3.0 
at larger values of $\sigma$, and then starts to fall for $\sigma$ greater
than about 400. Fitting the data to a function of the form
\mym{D(\sigma)=a+\frac{b}{c+\sigma}} we obtain
\begin{equation}
D(\sigma) = 3.03 - \frac{10.51}{17.87+\sigma}
\end{equation}
Fitting the data to \mym{D(\sigma)=a+b e^{-c\sigma}}, as suggested in \cite{C6}, we obtain
\begin{equation}
D(\sigma) = 3.19 - 0.97e^{-0.013\sigma}
\end{equation}
From this, we get \mym{D(0)=2.22} and \mym{D(\infty)=3.19}, which agrees well with
\mym{D(0)=2.12} and \mym{D(\infty)=2.98} reported in \cite{C6}.

The smooth reduction of the dimension of the spacetime at shorter scales is 
one of the most interesting features of the CDT model. This is an indication 
of the highly non-classical behavior at small scales. The reduction in 
dimension for larger $\sigma$ is due to the finite (and rather small) volume of 
the spacetimes being considered.

\begin{figure}[htp]
\centering
\includegraphics[scale=0.5]{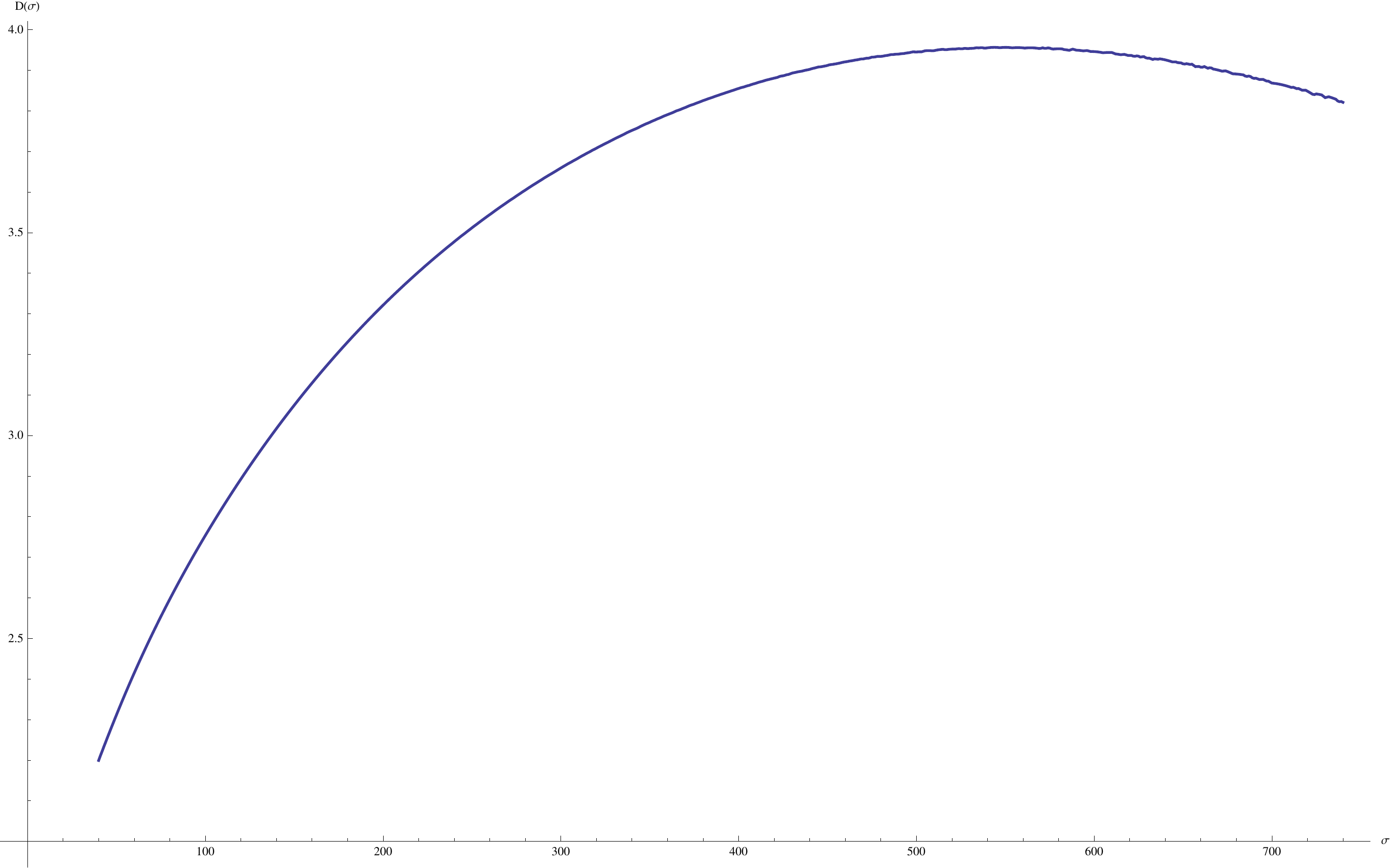}
\caption{Spectral dimension for a (3+1) spacetime vs $\sigma$, $N_4=80000,\kappa_0=3.6,\Delta=0.6$}
\label{fig:3p1-spectral-dimension}
\end{figure}

The spectral dimension vs step size plot for a (3+1)-dimensional spacetime
is shown in figure \ref{fig:3p1-spectral-dimension}.
The spectral dimension in this case ranges from a value of about 1.8 at short 
scales up to a value of about 4.0 at larger values of $\sigma$, and then starts 
to fall for $\sigma$ greater than about 600, with the functional form being
given by:
\begin{equation}
D(\sigma) = 4.43 - \frac{375}{141+\sigma}
\end{equation}
This is to be compared with the
\begin{equation}
D(\sigma) = 4.02 - \frac{119}{54+\sigma}
\end{equation}
reported in \cite{C2}.

\subsection{Spectral Dimension of the Spatial Slices}
The spectral dimension of a spatial slice can be computed using 
a process identical to that used for determining the
spectral dimension of the full spacetime. We select the slice with
the maximum $(d-1)$-volume and initiate diffusion processes from a
randomly selected $(d-1)$-simplex on this slice.
 
\begin{figure}[htp]
\centering
\includegraphics[scale=0.5]{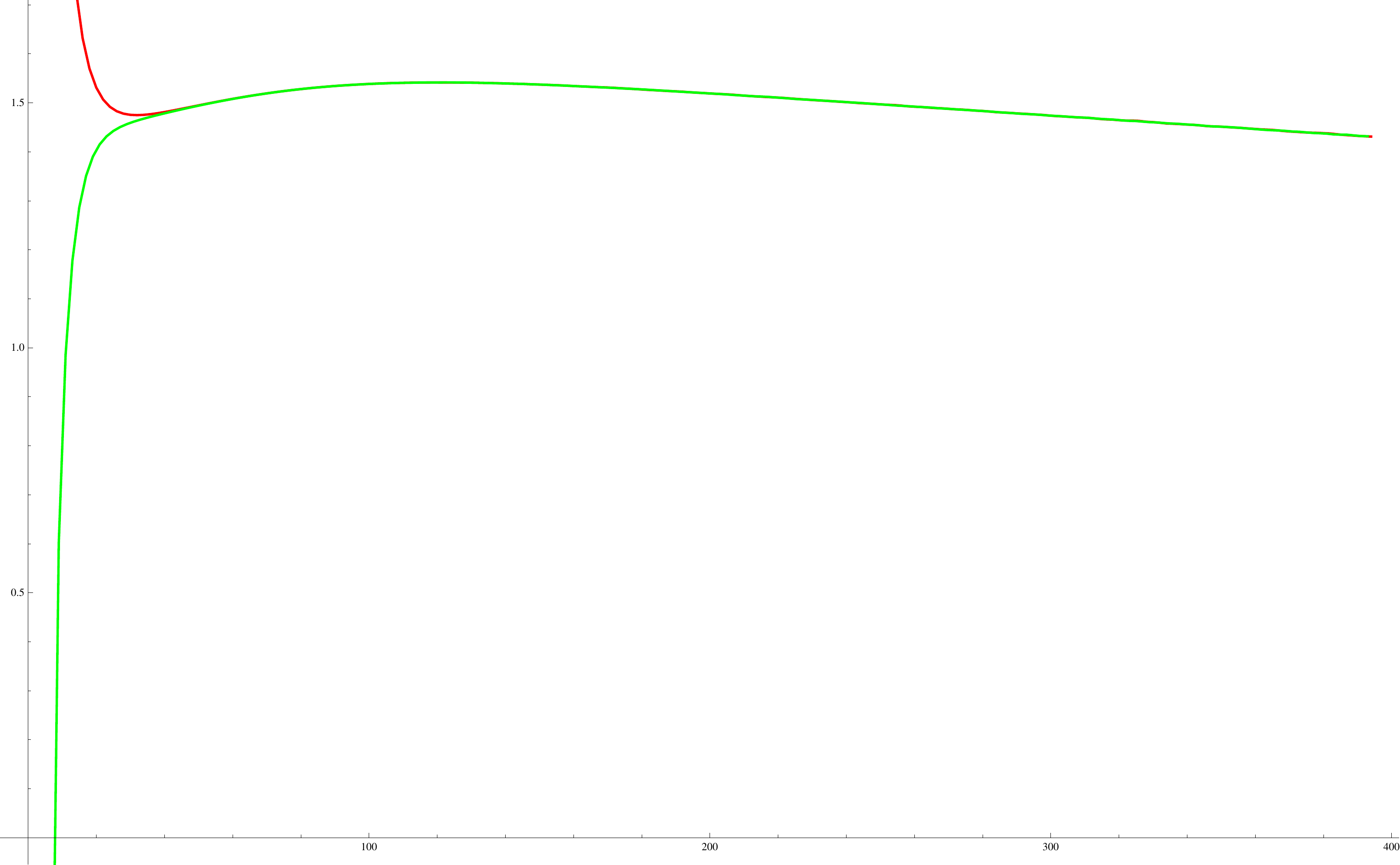}
\caption{Spectral dimension for the maximum volume spatial slice for a 3+1
spacetime}
\label{fig:3p1-sss}
\end{figure}

The spectral dimension of the maximum volume slice for a (3+1)-dimensional 
spacetime is show in figure \ref{fig:3p1-sss}. For short scales, 
up to $\sigma \approx 40$, the return probability differs considerably
for even and odd step sizes, and the plot bifurcates. Once the even and odd
step size spectral dimensions converge, $D(\sigma)$
is approximately 1.5, before a gradual reduction due to finite size
scaling. In \cite{C2}, a spatial slice spectral dimension of $1.56$ is 
reported. This is value is computed by calculating the average return
probability of all the spatial slices. The agreement of our results with
value reported in \cite{C2} indicates that each spatial slice has roughly
the same spectral dimension.   

\begin{figure}[htp]
\centering
\includegraphics[scale=0.5]{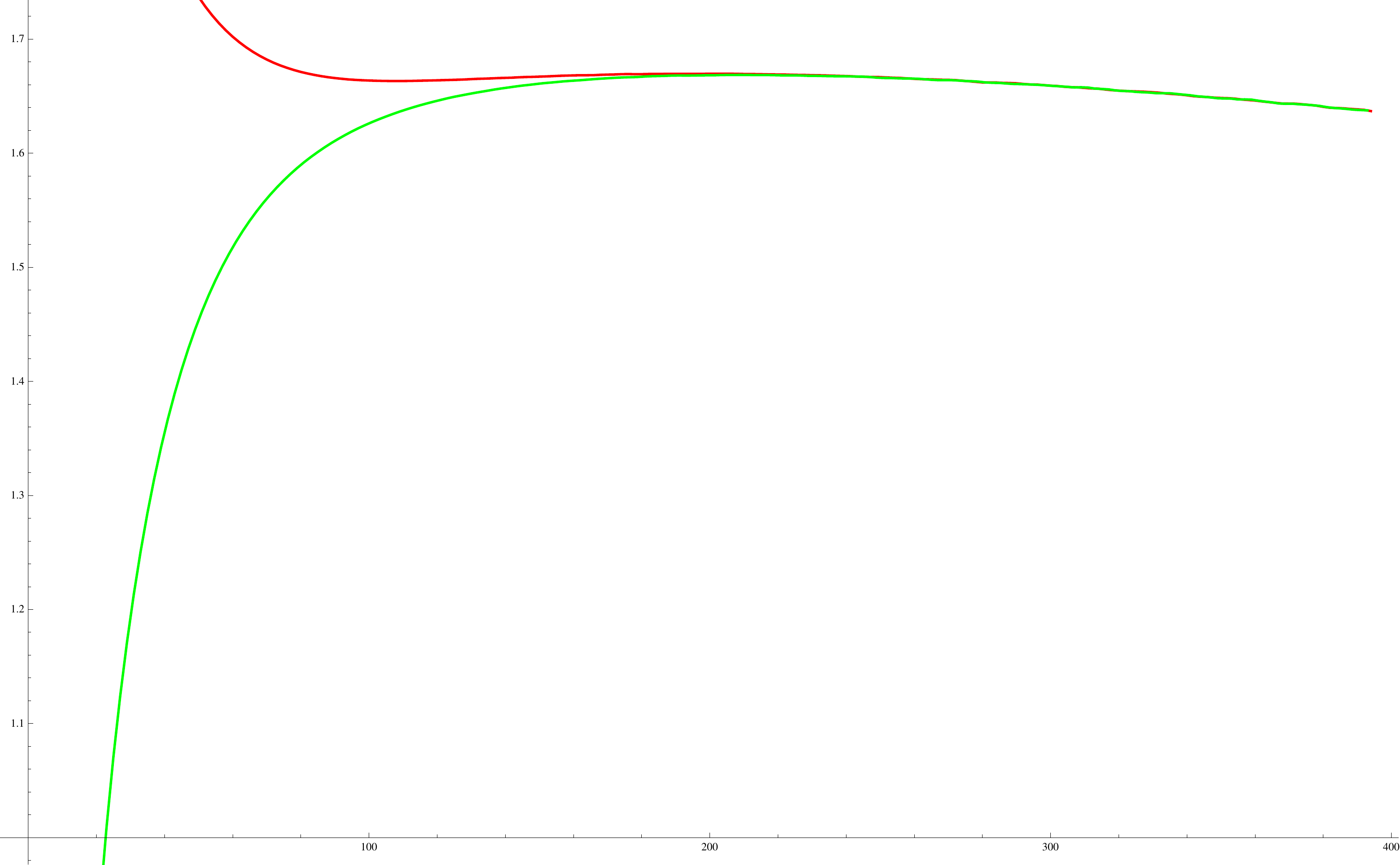}
\caption{Spectral dimension for the maximum volume spatial slice for a 2+1
spacetime}
\label{fig:2p1-sss}
\end{figure}

Figure \ref{fig:2p1-sss} shows a corresponding plot for (2+1)-dimensional 
spacetime. In this case, the even and odd $\sigma$ plots converge at a much
larger value of $\sigma \approx 150$. The spectral dimension value then 
levels off at about 1.65

\subsection*{Volume Profile}
At the beginning of this section, we discussed the emergence of an
extended phase, which we said was a solution to the classical Einstein 
equations. However, the only evidence we have presented so far are figures
\ref{fig:extended-phase} and \ref{fig:3p1-phase-C}, where the shape of the 
volume profile is indicative of such a solution. There is a rigorous, 
quantitative procedure, in which volume profiles of the spatial 
slices are fit to the Euclidean de Sitter solution, in 2+1 and 3+1 dimensions. 
This procedure is described in detail in \cite{C3}. We have been able to verify 
these results in 2+1 and 3+1 dimensions; our analysis will be presented in
\cite{Cooperman}.

\section{Conclusions}
The Causal Dynamical Triangulations approach to quantum gravity has many 
promising features, 
and our computer simulation has been able to successfully reproduce these results in 
2+1 and 3+1 dimensions. Most of these results were first reported by Ambj\o rn, 
Jurkiewicz, and Loll \cite{C0,C4,C1,C2}, and our simulation
provides the first completely independent verification. We observe the 
emergence of a well defined extended geometric phase. This geometric phase 
shows a smoothly changing spectral dimension that ranges from a value close to 
2.0 to a value larger than 3.0 for the (2+1)-dimensional spacetimes and from
about 2.0 to 4.0 for the (3+1)-dimensional spacetimes. The spatial slices exhibit
a non-classical behavior, as indicated by their spectral dimension plots.

Our successful reproduction of many of the results reported by Ambj\o rn, 
Jurkiewicz, and Loll, using a completely independent software implementation, 
should improve confidence in the CDT model. With the exception
of the value of the phase transition point in the (2+1)-dimensional model, our results are in good
agreement with those reported in \cite{C1,C2,C3}, and convince us of the overall validity of the
CDT approach.

Having successful reproduced the basic results of the CDT model, we are now
using our CDT implementation to explore some further interesting problems. These
include investigating the quantum fluctuations in (2+1)-dimensional CDTs and
comparing the results with canonical quantization approaches \cite{Sachs}, 
detailed analysis of the volume profiles in 2+1 and 3+1 dimensions 
\cite{Cooperman} and exploring Horava-Lifshitz gravity on the CDT lattice 
\cite{Zulkowski}. In the near future, we plan to release our implementation 
into the public domain.

\section*{Acknowledgements}
I would like to thank Professor Steve Carlip for his initial suggestion of CDT 
as a research topic, and for his guidance and patience for the duration of 
this project. I would like to thank the summer REU students, Jun Zhang, 
Masha Baryakhtar, David Kamensky and Christian Anderson, for their valuable 
contributions to the project. I would also like to thank Joshua Cooperman, 
Adam Getchell, Michael Sachs and Patrick Zulkowski for helpful discussions 
regarding CDT and for their willingness to work with poorly documented code 
written in an unfamiliar programming language.

This work was supported in part by DOE grant DE-FG02-91ER40674.

\end{document}